\documentclass[pra,12pt]{revtex4}
\usepackage{bm}
\usepackage{epsfig,amsbsy}
\begin{document}
\def\9{\rangle}
\def\1{\mbox{1\hskip-.25em l}}
\def\HD{{\cal H}_D}
\def\H{{\cal H}}
\def\D{{\cal D}}

\def\beq{\begin{equation}}
\def\eeq{\end{equation}}
\def\bea{\begin{eqnarray}}
\def\eea{\end{eqnarray}}
\def\half{\mbox{$1\over2$}}
\def\halfs{\scriptscriptstyle 1/2}
\def\halfb{\mbox{$\beta\over2$}}
\def\halft{\mbox{$\beta\over2$}}
\def\qtr{\mbox{$1\over4$}}
\def\ses{\mbox{$1\over16$}}
\def\rA{{\rm A}}
\def\rB{{\rm B}}
\def\bA{{\bf A}}
\def\bK{{\bf K}}
\def\bL{{\bf L}}
\def\bJ{{\bf J}}
\def\ba{{\bf a}}
\def\bk{{\bf k}}
\def\bu{{\bf u}}
\def\bv{{\bf v}}
\def\bz{{\bf z}}
\def\bn{{\bf n}}
\def\bx{{\bf x}}
\def\by{{\bf y}}
\def\bw{{\bf w}}
\def\br{{\bf r}}
\def\bs{{\bf s}}
\def\bb{{\bf b}}
\def\bt{{\bf t}}
\def\bu{{\bf u}}
\def\bv{{\bf c}}
\def\bM{{\bf M}}
\def\bmu{\boldsymbol{\mu}}
\def\ca{{\cal A}}
\def\cb{{\cal B}}
\def\cH{{\cal H}}
\def\cF{{\cal F}}
\def\cU{{\cal U}}
\def\cR{{\cal R}}

\def\hbb{{\hat{\bb}}}
\def\hbr{{\hat{\br}}}
\def\hbx{{\hat{\bx}}}
\def\hbz{{\hat{\bz}}}
\def\hby{{\hat{\by}}}
\def\hbn{{\hat{\bn}}}
\def\hbmm{{\hat{\bf m}}}
\def\hatt{{\hat{\beta}}}

\def\tal{{\tilde\beta}}
\def\teb{{\tilde\phi}}
\def\tap{{\tilde\varphi}}
\def\trho{{\tilde\rho}}

\def\brho{{\bar{\rho}}}
\def\bsigma{{\bar{\sigma}}}

\def\bel{\mbox{\boldmath $\ell$}}
\def\bl{\mbox{\boldmath $\ell$}}
\def\1{\mbox{1\hskip-.25em l}}
\def\6{\langle }
\def\9{\rangle }
\def\tr{{\rm Tr}}

\def\A{_{\!A}}
\newcommand\matrixx[1]{\left(\begin{matrix}#1\end{matrix}\right)}
\newcommand\smatrixx[1]{\left(\begin{smallmatrix}#1\end{smallmatrix}\right)}
\newcommand{\ket}[1]{ | \, #1  \rangle}
\newcommand{\bra}[1]{ \langle #1 \,  |}
\newcommand{\proj}[1]{\ket{#1}\bra{#1}}

\title{Quantum estimation of relative information}

\author{Netanel H. Lindner}

\email{lindner@tx.technion.ac.il}

\author{Petra F. Scudo}

\email{scudo@tx.technion.ac.il}  \affiliation{Department of
Physics, Technion---Israel Institute of Technology, 32000 Haifa,
Israel}

\author{Dagmar Bru{\ss}}
\email{bruss@thphy.uni-duesseldorf.de}\affiliation{Institut f\"ur Theoretische Physik III,
Heinrich-Heine-Universit\"at D\"usseldorf, D-40225 D\"usseldorf,
Germany}


\begin{center}

\textit{This work is dedicated to the memory of Asher Peres,
teacher and friend, whom we shall always greatly miss.}\\

\end{center}

\begin{abstract}

We derive optimal schemes for preparation and estimation of
relational degrees of freedom between two quantum systems. We
specifically analyze the case of rotation parameters representing
relative angles between elements of the $SU(2)$ symmetry group.
Our estimation procedure does not assume prior knowledge of the
absolute spatial orientation of the systems and as such does not
require information on the underlying classical reference frame in
which the states are prepared.

\end{abstract}


\maketitle

\section{Introduction}

Estimating the state of a given quantum system is a fundamental
primitive of many quantum information tasks. This problem is
usually translated to the estimation of the value of a physical
parameter describing specific properties of the preparation
procedure. In many instances
\cite{mp,gp,scudo1,scudo2,bagan1,bagan2,lindner,dariano,holevo1}
global parameters of the state space define a natural scheme for
encoding quantum information. The global parameters describe
collective degrees of freedom of a system with respect to the
external environment and are often related to an overall symmetry
transformation of the state.

However, encoding information into global degrees of freedom may
be often problematic, due to lack of knowledge of the reference
frame with respect to which they were prepared, or due to
collective decoherence by which they are affected. Encoding
information into relative degrees of freedom, possible whenever a
quantum system is decomposable into parts, can overcome many of
the difficulties encountered in these situations. Such an encoding
scheme has been demonstrated experimentally \cite{banaszek,kwiat}
and can be applied to quantum computation \cite{zanardi},
communication \cite{bourennane,commnorf} and cryptography
\cite{walton, boileau}.

The aim of this work is to develop efficient preparations and
measurement schemes for the relative parameters describing
symmetries between different components of a system. We note that
such measurements can induce relative relations when previously
absent, as in the case of the relative phase between two Fock
states or the relative position between two momentum eigenstates
\cite{hugo}.

In this paper we specifically confront the task of efficient
estimation of relative rotation angles between two representation
vectors of the $SU(2)$ symmetry group. This problem was first
addressed by Bartlett {\it et. al.} \cite{terry}, who explicitly
worked out the estimation of $SU(2)$ rotation angles between two
spin coherent states. In this article  we proceed along the lines
of earlier work
that Asher Peres started together with us, and propose an
extension of the previous methods. Our approach is based on an
optimization of the quantum states used in such protocols with
respect to some average measure of success of the estimation task,
which we shall refer to as the fidelity. The key to this problem
lies in the decomposition of both the signal and the measurement
elements in irreducible components, invariant under global
rotation transformations.

In the following section we discuss the general mathematical
structure of the problem. In section~\ref{Relative rotations of
$j_2= 1/2$} we derive the optimal measurement for the case in
which one system is comprised of two spin-$1/2$, and the other of
one spin-$1/2$. We find that preparing
 two spin-$1/2$
parallel to each other leads to a marginally higher fidelity than
the antiparallel case;
 then we
determine an optimal preparation procedure which gives a higher
fidelity then the ones achieved by the above preparations. This is
in contrast to the known results for transmitting a spatial
direction
\cite{gp}.
We then proceed  in section~\ref{higher values of $j_2$} by
replacing  the single spin state of the second system with a
spin-$j$ coherent state and determine the optimal preparations for
the cases of anti-parallel and parallel spins, which now yield
nearly the same fidelity for any value of $j$. We then study the
quantum/classical correspondence by considering the limit in which
$j$ becomes very large. In section \ref{Quantum-classical
correspondence} it is shown that in this limit the problem reduces
to estimating the first state with respect to a classical
reference direction. Our results establish the correspondence
between relative degrees of freedom in quantized systems and
collective degrees of freedom defined with respect to a classical
reference frame.

\section{Estimation of relative parameters}
\label{section:estimation}
\subsection{Formulation of the general problem and basic notations}
\label{section:definition}

We begin this section by formulating the problem for a general
symmetry group and introducing the basic notation that is useful
for our general scheme. In the following, let $G$ denote a
symmetry group, compact or finite, that describes the global
properties of the system through its action on a set of parameters
$T$. We shall consider $G = SU(2)$ acting on quantum spin states,
parametrized by the set of rotation angles. The possible states of
the system are pure states $\Psi (\Theta), \Theta \in T$ in a
$d$--dimensional Hilbert space $\H$ carrying a unitary
representation $\{U(g), g \in G\}$ of $G$. To introduce relative
symmetry transformations, we assume that the representation space
$\H$ is a tensor product of two components $\H = \H_1 \otimes
\H_2$ of dimensions $d_1, d_2$
respectively. The representation of $G$ on $\H$ is decomposed into
the product $\{U_1(g) \otimes U_2(g)\}$ on $\H_1 \otimes \H_2$.
This product representation of $G \times G$, in which each
component is transformed by the same element $g \in G$, is
isomorphic to $G$ itself. We introduce another set, $t$, of
parameters $\theta$, to describe a relative symmetry between the
two components, represented by a group of transformations
$\tilde{G}$, which can be the same $G$ as before or a subgroup of
the latter. $\tilde{G}$ refers to a symmetry property of one of
the subsystems (say 2) with respect to the other (say 1), such as,
in our case, a relative rotation angle. We call $U_2(h)$ its
representation operators on the space $\H_2$. We shall restrict
ourselves to the case where each of the subsystems is prepared in
a pure state, although the formalism can easily be extended to
mixed states. The total state on $\H$ can be written in terms of
the two sets of parameters as $\Psi(\Theta, \theta)$. Its
transformation under a global operation is
\begin{equation}
U(g) \Psi(\Theta, \theta) \equiv U_1(g) \otimes U_2(g)\Psi(\Theta,
\theta)= \Psi(g\, \Theta, \theta).
\label{eq: global operation}
\end{equation}

The objective of the construction is to define an efficient
estimation procedure for the relative parameters $\theta$,
overlooking the information carried by the global parameters
$\Theta$. Note that we might be interested only in estimating a
subset of the relative parameters (which will be the case in the
following sections). In order to quantify the efficiency of our
estimation procedure, we choose a utility function $f(\mu,\theta)$
which measures the deviation of the estimated parameters $\mu$
from their true values $\theta$. We consider only utility
functions which are invariant under global rotations. The
measurement apparatus, represented by the POVM $\{E_\mu\}$, should
be constructed such that it maximizes the average fidelity,
denoted by $F$, and given by
\beq
F\left\{E_{\mu}\right\}=\sum_{\mu}\int d\theta d\Theta
P(\Theta,\theta)\tr\left[\rho(\Theta,\theta)E_{\mu}\right]f(\mu,\theta)
\label{eq: def fidelity},
\eeq
  where $P(\Theta,\theta)$ is a prior
probability distribution over the global and relative  parameters,
and $\rho(\Theta,\theta)=\ket{\Psi(\Theta,\theta)}
\bra{\Psi(\Theta,\theta)}$.
As noted in \cite{terry}, we can assume that the global and
relative parameters are independent random variables, and that the
global parameter is uniformly distributed on its domain of
definition, {\it i.e.}, \beq P(\Theta,\theta)d\theta
d\Theta=p(\theta)d\theta d\Theta. \eeq Now, using the definition
for the global transformation~(\ref{eq: global operation}) and the
properties of the trace, we can write
 \beq
F\left\{E_{\mu}\right\}=\sum_{\mu}\int d\theta
p(\theta)\tr\left[\brho(\theta)E_{\mu}\right]f(\mu,\theta),
\label{eq: fidelity average rho}
\eeq where

\beq \brho(\theta)=\int\,
dg\,U^{\dagger}(g)\rho(\Theta,\theta)U(g).
\label{groupav} \eeq
In the last equation, $dg$ is the invariant measure for the group
$G$. As a consequence of equations (\ref{eq: fidelity average
rho}) and (\ref{groupav}), we need only to consider a reduced form
of the input state which is manifestly invariant under global
transformations. Schur's lemma \cite{sternberg} then assures that
the input state is block diagonal in the irreducible
representations of $G$,
\beq \brho(\theta)=\sum_{\oplus J}
\rho^{(J)}(\theta).
\eeq
The above considerations also have implications  on the form of
the optimal measurement. In fact, using the global invariance of
$\brho$, we have
 \bea
F\left\{E_{\mu}\right\}&=&F\left\{U(g)E_{\mu}U^{\dagger}(g)\right\}\nonumber\\
&=&F\left\{\int dg\,U(g)E_{\mu}U^{\dagger}(g)\right\}. \nonumber
\label{eq: invariant fidelity} \eea
This relation implies that when searching for the optimal
estimation procedure it suffices to consider POVMs which are
invariant under a global transformation
\begin{equation}
U(g) E_{\mu} U^{\dagger}(g) = E_{\mu}, \forall g \in G.
\label{inv}
\end{equation}
Equation (\ref{inv}) is by itself a strong prerequisite on the
structure of the POVM elements. Indeed, if combined with Schur's
lemma, it implies that whenever the total Hilbert space $\H$ can
be decomposed into a direct sum of irreducible representations
under the global transformation, the optimal POVM elements
labelling the different outcomes have the simple form
\begin{equation}
E_{\mu}=\sum_J p_{\mu, J} E^{J}_{\mu},
\label{eq: invariant POVMs}
\end{equation}
where each of the operators $E^{J}_{\mu}$ has support only on the
representation labelled by $J$ and $\sum_{\mu}p_{\mu,
J}E^{J}_{\mu}=\1_J$. Here the symbol $\1_J$ denotes the identity
operator in the $J$ subspace. The search for optimal POVMs may be
further restricted to the case in which all elements have support
on only one representation space, due to the linearity of the
fidelity functional (given a POVM of the form (\ref{eq: invariant
POVMs}), the POVM with elements $E_{\mu,J}\equiv p_{\mu,
J}E^{J}_{\mu}$ yields the same fidelity as the original one by
linearity of the trace and obeys the restriction that each element
has support on only one representation).

In the following sections, the general considerations outlined
above will be applied to the problem of transmitting relative
rotation angles of the $SU(2)$ group.

\subsection{Estimation of a relative angle between spin coherent states}
\label{section: Estimation of relative rotation angles}

In estimating relative, as opposed to absolute, rotation angles,
we assume no prior knowledge of the overall orientation of the
classical frame in which the system is defined. Following the
notation introduced above, the problem may be illustrated as
follows.

Imagine that, with no prior knowledge on the absolute spatial
orientation of two observers, Alice and Bob, we were requested to
estimate the angle $\beta$ between two unit vectors $\hbn_1$ and
$\hbn_2$, each chosen by one of them, by measuring a pair of
$SU(2)$ spin states prepared in the corresponding reference frame
of each observer. This task is in general possible owing to the
fact that states belonging to an $SU(2)$ representation space can
be used as intrinsic direction indicators
\cite{mp,gp,scudo1,scudo2,bagan1,bagan2,lindner,dariano,holevo1}
and therefore it makes sense to consider relative angles between
them.

The simplest way to achieve the task would be to consider two
$SU(2)$ coherent states, corresponding, say, to spins $j_1, j_2$,
\begin{equation}
\hbn_1\cdot\bJ_1|\psi_1\9=j_1|\psi_1\9,\qquad
\hbn_2\cdot\bJ_2|\psi_2\9=j_2|\psi_2\9.
\label{eq:coherent states}
\end{equation}
Without loss of generality, we can assume that Alice and Bob
choose the $z$-axis of their reference frame. A state denoted by
$\ket{\psi_2}$ in Bob's frame is written in Alice's reference
frame as $ U_2^z(\alpha)U_2^{\,y}(\beta)U_2^z(\gamma)\,
\ket{\psi_2}$, where
$U_k^x(\alpha)=e^{\,i\,\bJ_{k}\cdot\bx\,\alpha}$, and similarly
for the other directions. The angles $\alpha,\beta,\gamma$ are the
three Euler angles relating Alice's reference frame to the one of
Bob. Note that the angle $\beta\in [0,\pi]$ is also the angle
between $\hbn_1$ and $\hbn_2$. We introduce a global reference
frame which is rotated with respect to Alice's frame by an angle
$\alpha$ around the $z$-axis. In this frame the composite state is
given by
\beq
\ket{\Psi(\alpha,\beta,\gamma)}=U_1^z(\alpha)\, \ket{\psi_1}\otimes
U_2^z(\beta)U_2^z(\gamma)\, \ket{\psi_2}. \label{eq:general
states}
\eeq
For spin coherent states, as in Eq.~(\ref{eq:coherent states}),
the above equation is simplified to
\begin{equation}
\ket{\Psi(\beta)}= |j_1, m_1=j_1\rangle \otimes U_2^{\,y}(\beta) |j_2,
m_2=j_2\rangle,
\label{eq:example su(2) coherent states}
\end{equation}
up to an overall phase.  Notice that in Eq.~(\ref{eq:general
states}),(\ref{eq:example su(2) coherent states}) we have
implicitly specified the global parameter $\Theta$, which we shall
omit from now on from our notation. So far we have a set of three
relative parameters, $\theta=\{\alpha,\beta,\gamma\}$, with a
joint probability distribution given by the Haar-measure
\beq
p(\alpha,\beta,\gamma)=\frac{1}{8\pi^2}\sin\beta, \label{eq:prior
distribution relative}
\eeq
which corresponds to a random orientation of Alice's and Bob's
reference frames. Since the party making the measurement (Bob) is
not interested in estimating $\alpha,\gamma \in [0, 2\pi]$, these
parameters are averaged out by integrating over their range. We
are then left with the probability distribution
$p(\beta)=\sin\beta/2$, which corresponds to the probability
density of the angle between two random unit vectors in three
dimensions.

Let us denote by $\H_1, \H_2$ the Hilbert spaces of the systems
prepared by Alice and Bob respectively, carrying the $SU(2)$
representations $j_1$ and $j_2$. The composite Hilbert space
$\H=\H_1\otimes \H_2$ carries a diagonal product representation of
$SU(2)\otimes SU(2)$, $U^{j_1}(g)\otimes U^{j_2}(g), g \in SU(2)$,
which corresponds to a {\it global} symmetry operation labelled by
the parameter $g$, acting identically on the two subspaces. This
representation may be reduced as $\sum_{\oplus J=|j_1 -
j_2|}^{j_1+j_2} \H_J$, where each component has multiplicity one.
Invariance of the measurement operators under a global rotation
(as explained in the previous section) reduces the signal state to
the form
\begin{equation}
\brho(\beta)= \sum_J p_J(\beta) \Pi_J,
\end{equation}
where $\Pi_J$ are projectors on the representations $J$. Since
each representation of the global rotation (specified above by
$J$) appears {\it only once} in the signal state, the measurement
process amounts to estimating a probability distribution over the
relative angle $\beta$. The scenario described above, where each
of the parties prepares a spin coherent state, is the one examined
in \cite{terry}.

\section{General Scheme}
\label{general scheme}

In the procedure outlined in the last section, the two parties,
Alice and Bob, use spin coherent states in order to indicate their
chosen direction. However, for the task considered here this is
not the optimal preparation. It is known, in fact, that optimal
$SU(2)$ direction indicators exploit entanglement between
components belonging to different irreducible representations
\cite{scudo1,bagan1,bagan2,scudo2,dariano}. This suggests to
consider the following general encoding procedure. Let
\begin{equation}
\ket{\Phi}= \sum_{j_1=0}^{j_{max}}\;\sum_{m_1=-j_1}^{j_1} a^{j_1}_{m_1} |j_1 \, m_1\rangle
\label{s1}
\end{equation}
be a generic state in $\cH_1$. By choosing a unit vector $\hbn_1$,
Alice would prepare the state $U(\hbn_1)\ket{\psi_1}$, \beq
\ket{\psi_1}=U(\hbn_1)\ket{\Phi}, \label{eq:alices state} \eeq where
$U(\hbn_1)$ is a unitary operator corresponding to the rotation
which carries Alice's $\hbz$-axis onto $\hbn_1$. Our goal is to
find the optimal state $\ket{\Phi}$ for the case in which Bob
indicates his direction $\hbn_2$ with a coherent state
$\ket{\psi_2}$ satisfying
\begin{equation}
\hbn_2\cdot\bJ_2\ket{\psi_2}=j_2\ket{\psi_2}.
\label{s2}
\end{equation}
Note that equations (\ref{eq:alices state}) and (\ref{s2}) are
written in Alice's and Bob's reference frames, respectively. In a
global reference frame, specified as in the last section, the
total state is given by
\bea
\ket{\Psi (\alpha,\beta,\gamma)}&=& U_1^z(\alpha\!)\,\ket{\psi_1} \otimes U^y_2(\beta)U_2^z(\gamma\!)
\,\ket{\psi_2}\nonumber\\
&=& \sum_{j_1 \, m_1} a^{j_1}_{m_1}e^{i\, m_1\alpha} |j_1 \,
m_1\rangle \otimes U^{y}_2(\beta) e^{i\, j_2\gamma}|j_2 \,
j_2\rangle. \label{psi}
\eea
As before, the state $\ket{\Psi}$ is expressed in terms of an
arbitrary orientation of the global reference frame, so that the
parameter $\Theta$ may be omitted. However, one can see that the
angles $\alpha,\gamma$ now induce relative phases between the
different components of the state. Since our protocol does not
deal with the estimation of $\alpha$ and $\gamma$ (only the angle
$\beta$ is considered), we will average over them in the
expression of the fidelity
\beq F=\sum_{\mu}\int
d_{\alpha\beta\gamma}\,dg
\tr\left[U(g)\rho(\alpha,\beta,\gamma)U^{\dag}(g)E_{\mu}\right]f(\mu,\beta),
\eeq
where we denoted $d_{\alpha\beta\gamma}\equiv
1/8\pi^2\sin(\beta)\,d\alpha \,d\beta\, d\gamma$. Integrating over
$\alpha$ and $\gamma$, we have \beq \frac{1}{4\pi^2}\int d\alpha\,
d\gamma\;\rho(\alpha,\beta,\gamma)=\sum_{m_1}c_{m_1}\rho_{m_1}\otimes
U^{y}_2(\beta)|j_2\, j_2\9\6j_2\, j_2|U^{y\dag}_2(\beta), \eeq
with \beq \rho_{m_1}=\ket{\psi_{m_1}}\bra{\psi_{m_1}}, \qquad
\ket{\psi_{m_1}}=\sum_{j_1=m_1}^{j_{max}}a_{m_1}^{j_1}|j_1,m_1\9, \eeq
and $c_{m_1}$ given by \beq c_{m_1}=\sum_{j_1}
|a^{j_1}_{m_1}|^2,\qquad \sum_{m_1}c_{m_1}=1. \eeq
 In the following we
will search for the optimal generic state $\ket{\Phi}$. Note that
after integrating over $\alpha$,\, $\gamma$ we get a convex
combination of states with different $m_1$. Thus the fidelity will
contain a linear combination of contributions from the different
$m_1$ sectors,
\beq
 F=\sum_{m_1}c_{m_1}F(m_1)\leq\max_{m_1}F(m_1).
\eeq
From the above discussion it is clear \cite{thanks} that the
optimal generic state can be taken with $m_1$ fixed, {\it i.e,} of
the form
\beq
\ket{\Phi_{m_1}}=\sum_{j_1=m_1}^{j_{max}}a_{m_1}^{j_1}|j_1
m_1\9.
\eeq
In the following we shall restrict ourselves to generic states of
this form. The rotation in Eq.~(\ref{psi}) by the relative angle
$\beta \in [0, \pi]$ is expressed in the standard Euler angle
notation as a rotation around the $y$-axis, given by the matrix
\begin{equation}
U^{y}_2 (\beta) |j_2 \, j_2\rangle\equiv\sum_m d^{j_2}_{m j_2}
(\beta)|j_2 \,m\rangle,
\end{equation}
where the $d^{j_2}_{m' m}(\beta)$ can be expressed using Jacobi
polynomials (see for example \cite{edmonds}). The superscript in
the above equation refers to the $(2j_2+1)$--dimensional
irreducible representation of spin $j_2$. The signal state
(\ref{psi}) is given explicitly (up to an overall phase) by
\begin{equation}
\ket{\Psi(\beta)}= \sum_{j_1} a^{j_1}_{m_1} \sum_{m_2}
d^{j}_{ m_2\,j_2}(\beta) |j_1 \, m_1, \, j_2 \, m_2\rangle,
\label{psi2}
\end{equation}
where $|j_1 \, m_1, \, j_2 \, m_2\rangle\equiv |j_1 \, m_1\rangle
\otimes |j_2 \, m_2\rangle$. Note that unlike the example
discussed in section \ref{section: Estimation of relative rotation
angles} and in \cite{terry}, we now exploit repeated irreducible
representations $J$, since each value of $j_1$ gives rise to a
series of total angular momentum $J=|j_1-j_2|, ..., (j_1+j_2)$.
The equivalent repetitions of the representation $J$ are labelled
by $j_1$, with $j_2$ being fixed. The state can be written in the
basis $(J \, M, j_1 \ j_2)$ using the unitary transformation
\begin{equation}
|j_1 \, m_1, \, j_2 \, m_2\rangle = \sum_{J \, M} C^{J \, M}_{j_1
m_1 \, j_2 m_2} |J \, M, \, j_1 \, j_2\rangle,
\end{equation}
with $C$ denoting the Clebsch--Gordan coefficients, and $M$
denoting the $z$ component of the total angular momentum.

We now need to compute the averaged state $\brho(\beta)$,
following Eq.~(\ref{groupav}). A corollary to Schur's lemma
\cite{sternberg} then states that, for irreducible representations
$\sigma, \tau$ of a group $G$, a group--averaged operator
satisfies
\begin{equation}
\int dg\,U^{(\sigma)}(g) \,A\, U^{(\tau)}(g)^{\dagger} =
\frac{\delta_{\sigma, \tau} \tr_{\sigma}(A) \; \1}{d_\sigma},
\label{trace}
\end{equation}
where $\delta_{\sigma, \tau}$ is a Kronecker delta over the
inequivalent representations and the trace is computed over the
$d_\sigma$-dimensional space of the irreducible representation
$\sigma$. The above corollary is applied to obtain the invariant
reduced density operator $\bar{\rho}(\beta)$. Remembering that the
global rotation operator $U(\Omega)$ is a direct sum of operators
$U(\Omega)=\oplus_J U^{(J)}(\Omega)$, we see from
Eq.~(\ref{trace}) that $\brho(\beta)$ is also block diagonal in
the representations $J$,
\begin{eqnarray}
\brho(\beta)&=&\int U(\Omega) \rho(\beta) U^{\dagger}(\Omega) d\Omega \nonumber\\
&=& \sum_{\oplus J}
\brho^{(J)}(\beta),
\end{eqnarray}
where the operators $\rho^{(J)}(\beta)$ bear the indices $j_1,
j_1'$ and are given by \beq \brho^{(J)}(\beta)_{j_1,\;j_1'}=\sum_M
\6 J,M,j_1'| \rho(\beta)|J,M,j_1\9. \eeq Note that by averaging
over the global rotation one does not diagonalize the operator
$\brho(\beta)$ with respect to the additional quantum number
$j_1$. The invariant signal state $\brho(\beta)$ is block diagonal
over the irreducible representations with off diagonal elements
across the repeated ones
\begin{widetext}
\beq \brho(\beta)=\sum_{\oplus J} \sum_{j_1 j'_1} \frac{1}{(2J+1)}
\sum_{M  m_2} a^{j_1}_{m_1} a ^{* j'_1}_{m_1} \left(d^{j_2}_{j_2
\, m_2}(\beta)\right)^2 \, C^{J\, M}_{j_1 m_1 \, j_2 m_2}\,C^{J \,
M}_{j'_1 m_1 \, j_2 m_2} |J, \, j_1 \rangle \langle J, \,j'_1|.
\label{rf} \eeq
Remember that $m_1$ is fixed.

\end{widetext}

Note that the state $\brho(\beta)$ does not, even implicitly,
depend on the orientation of the reference frame of the
measurement apparatus.  We will now use $\brho(\beta)$ to
determine the state $\ket{\psi_1}$ which will enable optimal
estimation of $\beta$. To this end, we fix a convenient figure of
merit as measure of the discrepancy between the estimated and the
given value of $\beta \in [0, \pi]$, namely the quadratic utility
function $f(\mu, \beta)=\cos^2((\mu - \beta)/2)$, where $\mu$ is
the estimated value of the parameter. The choice of the utility
function is not unique, and a different choice might lead to
different optimal states and POVMs. However, the optimization
procedure, as described below, is independent of this choice. The
above utility function has the advantage of having been broadly
used throughout earlier literature
\cite{bagan2,bagan1,scudo1,scudo2,lindner,dariano,mp,gp}.

We denote the POVM elements by $\{E_{\mu}\}$, with $\mu \in [0,
 \pi]$, and $\sum_{\mu}
 E_{\mu} = 1$. The average fidelity with respect to the given
figure of merit, integrating over all possible transmitted angles
$\beta$ and all possible inferred values $\mu$, is
\begin{equation}
 F\left[\{\mu\},\{E_{\mu}\}\right]= \sum_{\mu} \int {\rm
Tr}[\brho(\beta) E_{\mu}] \cos^2\left(\frac{\mu -
\beta}{2}\right) \sin\beta \, d\beta/2.
\label{fidelity}
 \end{equation}
 Note that the fidelity is a functional both of the set of
 estimates $\{\mu\}$ and of the POVM used for the estimation
procedure $\{E_{\mu}\}$, where to each estimate corresponds a

(single) POVM element. The probability of estimating $\mu$ for a
true angle $\beta$ is $\tr[\brho(\beta) E_{\mu}].$

The above expression can be rewritten by exchanging the order of
the integral with the trace (due to the linearity of the
integration and finiteness of the sum) and gives the fidelity in
the form
\begin{equation}
F\left[\{\mu\},\{E_{\mu}\}\right] = \sum_{\mu} \tr\{
A_{\mu}E_{\mu}\}.
\end{equation}
with
\beq
A_{\mu} \equiv \int \brho(\beta)
\cos^2((\mu - \beta)/2) \sin\beta \, d\beta/2.
\label{eq: def Amu}
\eeq
 Since $\brho(\beta)$ is
block diagonal, also $A_{\mu}$ can be written as a direct sum
$A_\mu=\sum_{\oplus J} A_{\mu}^J$.

In terms of the basis representation states of angular momenta
$(J; j_1, j_2)$, and using Eq.~(\ref{rf}), the operator $A_{\mu}$
can be explicitly written as
\begin{eqnarray}
A_{\mu}&=& \sum_{\oplus J} \sum_{j_1 j'_1} \frac{1}{2J+1}
\sum_{M  m_2}\Big\{ a^{j_1}_{m_1} \,a ^{* j_1'}_{m_1}
\,I^{j_2}_{m_2}(\mu)\phantom{\Big\}}\nonumber\\
 &\phantom{+}&\phantom{\big\{ }C^{J \, M}_{j_1 m_1 \, j_2 m_2}\,C^{J \,
M}_{j'_1 m_1 \, j_2 m_2} \Big\} |J, \, j_1 \rangle \langle J, \,
j'_1|,
\end{eqnarray}
where
\begin{equation}
I^{j_2}_{m_2}(\mu) \equiv \int \left(d^{j_2}_{j_2 \,
m_2}(\beta)\right)^2\cos^2((\mu - \beta)/2)
\sin\beta \, d\beta/2.
\label{int}
 \end{equation}
 This expression may be evaluated using the properties of the
 Wigner functions $d^j_{m \, m'}$ and their representation in terms
 of Jacobi polynomials \cite{edmonds}. The average fidelity can now
 be written as a sum of contributions from each subspace of given
 $J$ \beq F\left[\{\mu\},\{E_{\mu}\}\right]=\sum_J \sum_{\mu} {\rm
 Tr} (A^J_{\mu}E^J_{\mu}), \label{eq:av fidelity} \eeq where
 $\sum_{\mu}E^J_{\mu}=\1_J$. Given a set of estimates $\{\mu\}$,
the task of maximizing the expression $\sum_{\mu} {\rm Tr}
(A^J_{\mu}E^J_{\mu})$ is straightforward, at least numerically
 (for example, by using semidefinite programming \cite{yonina}).
However, in our approach, in order to maximize the average
fidelity~(\ref{eq:av fidelity}), we need to maximize over the set
 $\{\mu\}$, so that the maximal fidelity $F_{\rm max}$ will
actually be given by \beq F_{\rm
max}=\max_{\{\mu\}}\,\max_{\{E_{\mu}\}}F\left[\{\mu\},\{E_{\mu}\}\right].
\label{eq: optimization problem} \eeq

\subsection{The case of $j_2= 1/2$, $j_1\in\{0,1\}$}
\label{Relative rotations of $j_2= 1/2$}

 We will now solve the
optimization problem of Eq.~(\ref{eq: optimization problem}) for
the following case: the state $\ket{\psi_2}$ is a spin $1/2$
coherent state, while $\ket{\psi_1}$ is composed of two spin $1/2$
systems, so that $j_1\in\{0,1\}$. According to the discussion in
the previous section, we can restrict ourselves to two classes of
generic states $\ket{\Phi}$
\begin{equation}
\ket{\Phi_0}=  a |j_1=0 \,m_1= 0\9 + \sqrt{1-a^2}\, |j_1=1 \,m_1= 0\9,
\end{equation}
and
\begin{equation}
\ket{\Phi_1}=  |j_1=1 \, m_1=1\9
\end{equation}
Let us first discuss the case where $\ket{\Phi}=\ket{\Phi_1}$. In
this simple case, the state $\ket{\psi_1}$ is just a spin-$1$
coherent state or, viewed as composed of two spins, it is a
polarized state with parallel spins along the vector $\hbn_1$.
Coupling the representations of $\psi_1$ and $\psi_2$ gives
\begin{displaymath}
\textbf{1}\otimes\textbf{1/2}=\textbf{1/2}\oplus\textbf{3/2},
\end{displaymath}
and the operators $A_{\mu}^J$ are one-dimensional, making the
optimization trivial. In this case, the optimal measurement simply
consists in the projections onto the $J=1/2$ and $J=3/2$
subspaces, as there are no repeated representations. The fidelity
achieved with this state is $F_{\rm
max}\left[\textrm{parallel}\right]=0.90983$.

Next, we consider $\ket{\Phi}=\ket{\Phi_0}$. Coupling the
representations of $\psi_1$ and $\psi_2$ gives in this case
\begin{displaymath}
(\textbf{1}\oplus\textbf{0})\otimes\textbf{1/2}=\textbf{1/2}\oplus\textbf{1/2}\oplus\textbf{3/2},
\end{displaymath}
and therefore the density matrix $\brho(\beta)$ contains two
blocks of dimensions $2$ and $1$ corresponding to $J=1/2$ and
$J=3/2$, respectively. The operators $A^J_{\mu}$ are given by \bea
&&  A_{\mu}^{1/2}= \left(\begin{array}{cc}
\frac{a^2\ (4 + \pi\sin \mu)}{8} & \frac{a \sqrt{1-a^2} \cos \mu}{6 \sqrt{3}} \\
\frac{a \sqrt{1-a^2} \cos \mu}{6 \sqrt{3}} & \frac{6(2-a^2)+3\pi\left(1-a^2\right)\sin
\mu}{72}
\end{array} \right)\nonumber
\eea
 and
  \beq
  {\small
A_{\mu}^{3/2}=\frac{12\left(1-a^2\right)
+3\left(1-a^2\right)\sin\mu}{36}. } \label{amu 3/2}
\eeq
We are seeking the set $\mu$ and $E^J_{\mu}$ which maximizes the
mean fidelity~(\ref{eq:av fidelity}). Let us start with the
$J=3/2$ subspace. Since this subspace is one-dimensional, the
restriction to operators which are invariant under a global
rotation leaves us with one operator only, $E^{3/2}_{\mu_{3/2}}$,
which is the projection operator on the $J=3/2$ subspace. The
estimate $\mu_{3/2}$ which maximizes the corresponding expression
for $A^{3/2}_{\mu}$ in reference to Eq.~(\ref{amu 3/2}), is
obviously given by $\mu_{3/2}=\pi/2$.

Next, we consider the 2-dimensional subspace of $J=1/2$. Following
\cite{helstrom}, we define an operator $\Upsilon$ as \beq \Upsilon
= \sum_{\mu} A_{\mu}E_{\mu}. \label{eq:upsilon} \eeq For a set
$\{\mu\}$, a POVM  $\{E_{\mu}\}$ is optimal if and only if it
satisfies the following set of conditions \beq \Upsilon-A_{\mu}
\geq 0 \label{eq:solution} \eeq for each $\mu$ in the set of
estimates $\{\mu\}$, with the additional requirement that
$\Upsilon$ be hermitian. The inequality sign in
Eq.~(\ref{eq:solution}) means that the operator $\Upsilon-A_{\mu}$
must be positive semi-definite. The maximal fidelity will then be
given by
\beq
 F_{\rm max}=
\tr \,\Upsilon. \label{eq:fmax}
 \eeq

In order to see that equations~(\ref{eq:solution}) and
(\ref{eq:upsilon}) indeed lead to the maximization of the mean
fidelity, consider a different POVM $\{E_{\mu}'\}$, such that
\beq
\sum_{\mu} E_{\mu}' = \1. \label{eq:other POVM}
\eeq
The difference between the fidelity achieved with this POVM and
the one achieved with the optimal one is
\beq
F_{\rm max}-F'=\tr
\,
\sum_{\mu}(\Upsilon-A_{\mu})E_\mu',
\eeq
thanks to Eqs.~(\ref{eq:fmax}) and (\ref{eq:other POVM}). Now, if
$C$ and $D$ are positive semi-definite hermitian operators, then
they satisfy
\beq
\tr (CD) \geq 0. \eeq Setting $C= \Upsilon-A_{\mu}$ and
$D=E_{\mu}'$, we obtain \beq F_{\rm max}-F' \geq 0, \eeq as
desired.

Let us first maximize the average fidelity for only two estimates
$\mu_1$ and $\mu_2$, in correspondence to which we have the POVM
elements $E^{\halfs }_{\mu_1}+E^{\halfs }_{\mu_1}=\1_{J=1/2}$.
Then
\bea
\Upsilon-A^{\halfs }_{\mu_1}&=&A^{\halfs }_{\mu_1}E^{\halfs }_{\mu_1}+A^{\halfs }_{\mu_2}E^{\halfs }_{\mu_2}-A^{\halfs }_{\mu_1} \nonumber\\
&=& (A^{\halfs }_{\mu_2}-A^{\halfs }_{\mu_1})E^{\halfs }_{\mu_2}
\geq 0, \label{eq:proof1} \eea since $\Upsilon-A^{\halfs
}_{\mu_1}$ is non-negative if the POVM is optimal. Let us denote
by $\eta_i$ and $|\eta_i\9$ the eigenvalues and corresponding
eigenvectors of the operator $\Delta \equiv A^{\halfs
}_{\mu_2}-A^{\halfs }_{\mu_1}$. For each $|\eta_i\9$ we can write
\beq \6 \eta_i|(A^{\halfs}_{\mu_2}-A^{\halfs}_{\mu_1})
E^{\halfs}_{\mu_2}|\eta_i\9=\eta_i\6\eta_i|E^{\halfs}_{\mu_2}|\eta_i\9
\geq 0, \label{pe}\eeq using (\ref{eq:proof1}). If we assume that
$\eta_i$ is negative, then (\ref{pe}) gives
\begin{displaymath}
\6\eta_i|E^{\halfs }_{\mu_2}|\eta_i\9\leq 0;
\end{displaymath}
on the other hand, since $E^{\halfs }_{\mu_2}$ is positive
semi-definite, we must have
\begin{displaymath}
\6\eta_i|E^{\halfs}_{\mu_2}|\eta_i\9= 0, \qquad {\rm if} \;\;\eta_i<0,
\end{displaymath}
and similarly
\begin{displaymath}
\6\eta_i|E^{\halfs }_{\mu_1}|\eta_i\9= 0, \qquad {\rm if}\;\; \eta_i>0.
\end{displaymath}
Thus $E^{\halfs}_{\mu_2}$ projects onto the subspace spanned by
$|\eta_i\9$ with $\eta_i\leq 0$ and $E^{\halfs}_{\mu_1}$ projects
onto the subspace of positive eigenvalues of the operator
$\Delta$. The subspace with $\eta_i=0$ does not contribute to the
fidelity, so that the maximal fidelity is given by \bea F_{\rm
max}=\tr\,
\Upsilon = \tr \,A^{\halfs}_{\mu_1}+ \sum_{\eta_i\geq 0}\eta_i. \eea
Let us now assume that $\mu_1=\mu$ and $\mu_2=\pi-\mu$. We shall
see that this choice will lead to the optimal measurement for the
class of states under consideration. Indeed, we have
 \beq \Delta=
\left(\begin{array}{cc}
0 & \frac{a\,\sqrt{1-a^2}\cos \mu}{3\sqrt{3}}\\
 \frac{a\, \sqrt{1-a^2}\cos \mu}{3\sqrt{3}} & 0
\end{array}\right),
\eeq
with eigenvalues $\pm\frac{a\,\sqrt{1-a^2}\cos \mu}{3\sqrt{3}}$
and corresponding eigenvectors $|+\9=(1,1)^T$ and $|-\9=(1,-1)^T$.
The contribution to the fidelity from the $J=1/2$ subspace
$F^{\halfs}$ is now given by
\beq
F^{\halfs}(\mu)=\frac{a\,\sqrt{1-a^2}\cos
\mu}{3\sqrt{3}}+\tr\,A^{\halfs}_{\mu}.
 \label{Fmax}\eeq
At this point, in order to find the maximal mean fidelity under
our assumptions, it suffices to maximize the function
$F^{\halfs}(\mu)$. A simple calculation shows that the maximum is
attained for
 \beq
\nu=\tan^{-1}\left[3\sqrt{3}(1+2a^2)\pi\Big/(8a\sqrt{1-a^2})\right]
\label{eq:nu}. \eeq
It remains to check that indeed the choice
\beq \mu_1=\nu,\qquad \mu_2=\pi-\nu \label{eq:optimal guesses}
\eeq
leads to the maximal fidelity for the $J=1/2$ subspace. A proof of
this fact is provided by the following argument. Consider a
general set of estimates
$\{\bmu\}=\{\mu_1,\mu_2,\mu_3,...,\mu_n\}$, with a corresponding
set of POVM elements $E^{\halfs}_{\mu}$, and let
\beq
F^{\halfs}\left[\{\bmu\}\right]=\max_{\{E^{\halfs}_{\mu}\}}F^{\halfs}\left[\{\bmu\},\{E^{\halfs}_{\mu}\}\right]
\eeq
be the maximal fidelity achieved by this set. We would like to
show that by adding $\nu$ and $\pi-\nu$ to the set $\{\bmu\}$ the
mean fidelity can never decrease with respect to the optimal bound
and, at the same time, the bound is attained by these two values
alone. Let $\{\tilde{\bmu}\}$ denote the new set obtained by
adding $\nu$ and $\pi-\nu$, defined by Eq.~(\ref{eq:nu}), to
 the set $\{\bmu\}$. The first property, \beq
F^{\halfs}\left[\{\tilde{\bmu}\}\right]\geq
 F^{\halfs}\left[\{\bmu\}\right], \eeq simply follows from the fact
 that adding estimates to a given set can only increase the mean
fidelity. To complete the argument we still have to show that
\beq
F^{\halfs}\left[\{\tilde{\bmu}\}\right]=F^{\halfs}\left[\{\nu,\pi-\nu
\}\right]. \label{eq:last stage in proof} \eeq
The optimal measurement for $\{\nu,\pi-\nu \}$, following the
earlier discussion, is defined by the projectors on the negative
and positive eigenvectors of the operator
$A^{\halfs}_{\nu}-A^{\halfs}_{\pi-\nu}$, {\it i.e.},
\beq
E^{\halfs}_{\nu}=|-\9\6-|, \qquad E^{\halfs}_{\pi-\nu}=|+\9\6+|.
 \label{eq:POVM for two outcomes} \eeq
 Consider a POVM for the
set $\{\tilde{\bmu}\}$ which consists of the two operators in
(\ref{eq:POVM for two outcomes}), and of
\begin{displaymath}
  E^{\halfs}_{\mu}=0\qquad {\rm if}\qquad \mu\neq \nu, \pi-\nu.
\end{displaymath}
This POVM is optimal also for the set $\{\tilde{\bmu}\}$. To see
 this, we need to check whether the condition
\begin{displaymath}
\Upsilon-A^{\halfs}_{\mu} \geq 0
\end{displaymath}
holds for all $\mu \in \{\tilde{\bmu}\}$, with
\beq
\Upsilon=A^{\halfs}_{\nu}|-\9\6-|+A^{\halfs
}_{\pi-\nu}|+\9\6+|.
\eeq
Let us evaluate the entries of the operators
$\Upsilon-A^{\halfs}_{\mu}$ in the basis $|+\9,|-\9$. These are
given by
\bea
\6-|\Upsilon-A^{\halfs}_{\mu}|-\9=
\6-|A^{\halfs}_{\nu}|-\9 - \6-|A^{\halfs}_{\mu}|-\9 \nonumber\\
\6+|\Upsilon-A^{\halfs}_{\mu}|+\9=
\6-|A^{\halfs}_{\nu}|-\9 - \6+|A^{\halfs}_{\mu}|+\9 \nonumber\\
\6+|\Upsilon-A^{\halfs}_{\mu}|-\9= \6+|A^{\halfs}_{\nu}|-\9 -
\6+|A^{\halfs}_{\mu}|-\9, \label{eq:upsilon minus A} \eea
where we have used
$\6+|A^{\halfs}_{\pi-\nu}|+\9=\6-|A^{\halfs}_{\nu}|-\9$. The
eigenvalues $\lambda_1(\mu),\lambda_2(\mu)$ of the operator
$\Upsilon-A^{\halfs}_{\mu}$ can now be calculated from
(\ref{eq:upsilon minus A}), and their positivity can be verified
(at least numerically). The positivity of these eigenvalues (of
which we do not report here the explicit expression) implies then
(\ref{eq:last stage in proof}). For all input states
$\ket{\psi_1}$ discussed in this paper we have verified that the
POVM given in (\ref{eq:POVM for two outcomes}), with $\nu$ given
by (\ref{eq:nu}), is indeed optimal.

\begin{figure}[htbp]
\epsfxsize=.46\textwidth
\centerline{\epsffile{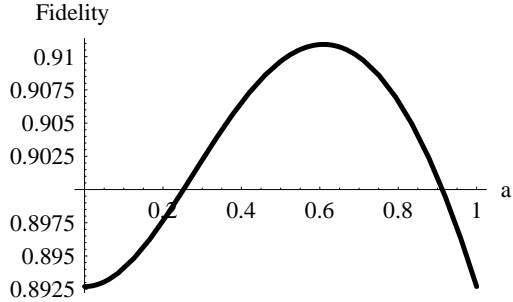}} \vspace*{-0.2cm}
\caption{\small{Fidelity as a function of the state parameter $a$.}}
\label{fig:one}
\end{figure}

From the above optimization procedure we see that the maximal
fidelity, as a function of the parameter $a$, is given by
\beq
 F_{\rm max}\left[m_1=0\right]=\frac{a\,\sqrt{1-a^2}\cos
\nu}{3\sqrt{3}}+\tr\,A^{\halfs}_{\nu}+\tr\, A^{3/2}_{\pi/2}
\label{optimal fidelity b=d},
 \eeq

  with $\nu$ given by
(\ref{eq:nu}). The fidelity as a function of the state parameter
$a$ is plotted in Fig.~\ref{fig:one}.
\\
The maximal fidelity is achieved in correspondence of the state
$\ket{\psi_{\rm opt}}$, by setting $a = 0.609$, and is $F_{\rm
max}\big[\psi_{\rm opt}\big] =0.91092$. For comparison, the
anti-parallel spin state
\begin{displaymath}
|\Phi_{\rm anti}\9=|\uparrow\downarrow\9=\frac{1}{\sqrt{2}}| 0 0\9
+\frac{1}{\sqrt{2}}| 1 0\9
 \end{displaymath}
leads to a fidelity of $F_{\rm
 max}\big[|\uparrow\downarrow\9\big]= 0.90982$, lower than the one
 obtained using the parallel spin state by only a factor $\sim 10^{-5}$.
 Similar results for parallel and anti-parallel spin states were
 obtained by N. Gisin and S. Iblisdir \cite{gi}.

To conclude this part, we compare the above results to earlier
results on quantum direction indicators, where a quantum system
carrying a representation of the rotation group is used to
transmit a spatial direction between two observes that do not
share a common reference frame. As shown in \cite{gp,scudo1}, if
the state of the quantum system is constructed from two spin
$1/2$, ({\it i.e.}, constrained to have maximal spin 1), encoding
the directional information into anti-parallel spins proves to be
the optimal strategy. Here we see instead that if the receiver is
interested only in the relative orientation of this state with
respect to another state, the anti-parallel spin state gives
nearly the same fidelity as the parallel one, which is well below
the optimum.

\subsection{Higher values of $j_2$}
\label{higher values of $j_2$}

The estimation of the relative orientation of two states can be
seen as a process in which the first party (Alice) encodes a
direction into a quantum state while the receiving party (Bob)
attempts to estimate the signal without having a classical
reference frame relative to which he can measure it. Therefore Bob
resorts to finding the relative orientation of the signal state
with respect to the orientation of some given state (say, a
coherent state of spin $j_2$), which serves as a quantum reference
frame. So far the value of $j_2$ has been kept equal to $1/2$. We
move on to consider what happens as we increase the value of
$j_2$. The limit $j_2\rightarrow \infty$ could be regarded as the
limit in which the quantum reference direction becomes a classical
one.

As before, we need to consider states with $m_1=0$ and $m_1=1$.
Different blocks of $\brho(\beta)$ are found by coupling the
representations of $|\psi_1\9$ and $|\psi_2\9$. For the $m_1=1$
sector, i.e. the case of parallel spins, $\brho(\beta)$ has three
one-dimensional blocks, since
\begin{displaymath}
\boldsymbol{1}\otimes\boldsymbol{j_2}=
(\boldsymbol{j_2-1})\oplus\boldsymbol{j_2}\oplus(\boldsymbol{j_2+1}).
\end{displaymath}
The optimal measurement is then given by projections on subspaces
of total angular momentum $J$.

For the $m_1=0$ sector we have
\begin{displaymath}
(\boldsymbol{0}\oplus\boldsymbol{1})\otimes\boldsymbol{j_2}=\boldsymbol{j_2}
\oplus(\boldsymbol{j_2-1})\oplus\boldsymbol{j_2}\oplus(\boldsymbol{j_2+1}),
\end{displaymath}
thus $\brho(\beta)$ and the operators $A_{\mu}$ have two
$1$-dimensional blocks and one $2$-dimensional block. The POVM
elements acting on the $J=j_2-1$ and $J=j_2+1$ subspaces are rank
one projectors onto these subspaces. In the $J=j_2$ subspace, an
optimization procedure similar to the one in Sec.~\ref{Relative
rotations of $j_2= 1/2$} needs to be done.

\begin{figure}[htbp]
\epsfxsize=.44\textwidth
\centerline{\epsffile{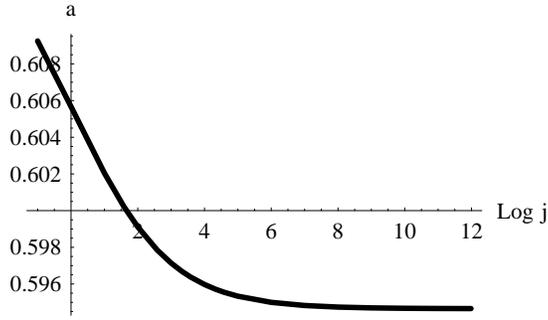}} \vspace*{-0.2cm}
\caption{\small{The optimal state parameter $a$ (vertical axis) with increasing values of $j_2$ (horizontal axis).}}
\label{fig:optimal state}
\end{figure}

Carrying out the optimization, we find that the optimal state
$\ket{\psi_{\rm opt}}$ belongs to the $m_1=0$ sector for all
values of $j_2$. The optimal state is not fixed but rather depends
on the value of $j_2$. The limit $j_2\rightarrow\infty$ yields the
optimal asymptotic state \beq \lim_{j_2\rightarrow
\infty}|\psi_{\rm opt}\9=a_\infty|00\9+\sqrt{1-a^2_\infty}|10\9,
\eeq with $a_\infty=0.595$. The dependence of the state $\ket{\psi_{\rm
opt}}$ on $j_2$ is plotted in Fig~\ref{fig:optimal state}.

\begin{figure}[htbp]
\epsfxsize=.44\textwidth
\centerline{\epsffile{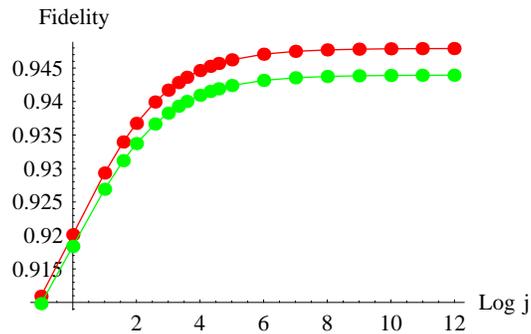}} \vspace*{-0.2cm}
\caption{\small{Fidelity achieved using the optimal state (red, upper
line), anti-parallel spins and parallel spins (green, lower line)
 as a function of $j_2$.}}
\label{fig:fidelity comparison}
\end{figure}

Comparing the fidelity achieved with the optimal state, as a
function of $j_2$, to the fidelity obtained using both the
anti-parallel and the parallel spin states, we get, quite
remarkably, almost the same fidelity for any value of $j_2$. The
parallel spins give a slightly higher fidelity, with the
difference (already very small for $j_2=1/2$) rapidly decreasing
for increasing values of $j_2$. This comparison is plotted in Fig
\ref{fig:fidelity comparison}, which shows that the plot for the
parallel and anti parallel spin states coincide.

Notice that although we expect the limit $j_2\rightarrow\infty$ to
be equivalent to measuring the state $\ket{\psi_1}$ against a
classical reference direction, the anti-parallel spin states do
not become optimal in this limit. This seems to be in
contradiction with the result of \cite{gp,scudo1}, who showed that
if only two spins are available, the optimal direction indicator
is provided by anti-parallel spins along that direction. The
resolution to this apparent contradiction relies on the fact that
the $j_2\rightarrow\infty$ limit considered here is not equivalent
to an estimation of a direction with respect to a classical
reference frame, as
\cite{bagan2,bagan1,scudo2,lindner,dariano,scudo1,mp,gp}, but
rather to that of an angle between the same vector, and, say, the
$z$-axis of such a frame. In the next section we will consider the
latter estimation task and show that it coincides with the limit
$j_2\rightarrow\infty$ discussed here, demonstrating that a
macroscopic spin can be treated as a classical reference
direction.

It is conceivable to consider a different estimation task in which
one observer, say Alice, indicates a direction in space using the
state $\ket{\psi_1}$, while Bob encodes his reference frame into
the state $\ket{\psi_2}$, and finally a third observer is
interested in the {\it orientation} of Alice's direction in Bob's
frame. This task, however, cannot be performed using a spin
coherent state $|\psi_2\9=|j_2,j_2\9$. It would be interesting to
see what state would encode Bob's frame in an optimal manner, and
whether taking the appropriate limit would reproduce previous
results for direction alignment.

\subsection{Quantum-classical correspondence}
\label{Quantum-classical correspondence}

Let us consider a scenario in which Alice prepares a state,
$\ket{\psi_1}$, in order to indicate a chosen direction, and Bob
is requested to estimate the angle between this direction and the
$z$-axis of his {\it classical reference frame}, which replaces
the quantum reference direction ($\ket{\psi_2}$) in the previous
study. We assume no knowledge of Alice's reference frame, and
without loss of generality we can assume that Alice chooses to
indicate her $z$-axis. If the transformation relating Alice's
frame to Bob's is parameterized by the Euler angles $\chi,\beta$
and $\phi$, then, in the latter reference frame, Alice's state is
given by $U(\chi,\beta,\phi)|\psi_1\9$. Bob's task is now to
estimate the angle $\beta$ between Alice's $z$-axis and his
$z$-axis. As before, given $\ket{\psi_1}$, we are seeking a POVM
that maximizes the fidelity
 \beq
F\left\{E_{\mu}\right\}=\sum_{\mu}\int d_{\chi\beta\phi}
\tr\left[\sigma_1(\chi,\beta,\phi)E_{\mu}\right]f(\mu,\beta),
\eeq
where $ d_{\chi\beta\phi}=\sin\beta d\beta d\phi d\chi
/8\pi^2$ is the invariant measure of the rotation group and
\begin{displaymath}
\sigma_1(\chi,\beta,\phi)=U(\chi,\beta,\phi)|\psi_1\9\6\psi_1|U(\chi,\beta,\phi)^{\dag}.
\end{displaymath}
 Since we are not interested in estimating the angles $\phi$ and $\chi$, we
can integrate over them and define an averaged density matrix
function of $\beta$ only, \beq \bsigma(\beta)=\frac{1}{4\pi^2}\int
d\phi \,d\chi \,\sigma_1(\chi,\beta,\phi). \eeq Let us analyze the
form of the density matrix $\bsigma(\beta)$. Writing
$\ket{\psi_1}$ as in Eq.~(\ref{s1}), and using the definition of
the rotation operator matrix elements \beq \6 j
m'|U(\chi,\beta,\phi)| j
m\9=e^{i\,m'\chi}\,d^{j}_{m'\,m}(\beta)e^{i\,m\phi}, \eeq we
derive the matrix elements of $\bsigma(\beta)$ as \beq \6 j'\,
m'|\bsigma(\beta)|j\, m\9=\sum_r\delta_{m'\,
m}a^{j\,'}_{r}a^{j*}_r d^{j\,'}_{m'\,r}(\beta) d^j_{m\,r}(\beta).
\eeq The matrix $\bsigma(\beta)$ is thus diagonal in the indices
$m$, but has off-diagonal elements with different values of $j$.
In this respect it is similar to the matrix $\brho(\beta)$ defined
in Eq.~(\ref{rf}), which was diagonal in the representation $J$
with off-diagonal elements between different values of~$j_1$.
Indeed, by taking the limit $j_2\rightarrow\infty$ in
Eq.~(\ref{rf}) and interchanging the indices $J\rightarrow m$, we
get the asymptotic equivalence between the matrices $\brho(\beta)$
and $\bsigma(\beta)$, {\it i.e.}, \beq \lim_{j_2\rightarrow
\infty}\6 J j_1|\brho(\beta)| J j_1^\prime\9=\6 j_1
m|\bsigma(\beta)| j_1'm\9. \eeq This result shows that the
fidelity achieved by relating any state $\ket{\psi_1}$ to a
classical reference direction is identical to the one achieved in
the limit $j_2\rightarrow\infty$ discussed in the previous section
(with the same state $\ket{\psi_1}$). Consequently, also the
optimal state will be identical in the two cases.

\section{Concluding remarks}

In this work we studied the problem of estimating relative
rotation angles of quantum signals. By considerations of global
symmetry invariance, we derived the general form of the signal
state and of the appropriate set of measurements and estimation
strategies. For special low-dimensional cases, explicit
optimization is carried out.

With these tools we have studied the state preparations which
maximize the average fidelity of the estimation procedure, and
compared these with the results found in the estimation of
absolute rotations. We have also discussed the asymptotic limit in
which one of the quantum states becomes a macroscopic spin. In
this limit, the resulting estimation task is identical to an
estimation of the orientation of a quantum state with respect to a
classical reference direction.

Many important questions remain open for investigation. A broader
extension of the problem would lead to the estimation of the
orientation of a quantum state relative to another one that
encodes a full reference frame (three axes), rather then a single
direction. In such a scenario one has to estimate two angles
(polar and azimutal). An even more elaborate framework is one in
which each quantum state encodes a full reference frame, and one
is interested in estimating the transformation that would align
these two reference frames. It would be interesting to compare the
optimal states found in all these cases with those found in prior
studies \cite{scudo2,scudo1,bagan1,bagan2,dariano}, when one of
the reference frames is classical. We would like, finally, to
emphasize that we did not include in our communication scheme the
possibility of the two parties  sharing a prior entangled state.
In this case, the encoding and the detection of relative
information could proceed via a covariant dense coding scheme,
which could increase the efficiency of the estimation.

\section*{Acknowledgements}

The Authors acknowledge interesting discussions with Daniel Terno,
Sofyan Iblisdir, Terry Rudolph and Giulio Chiribella. Special
thanks are due to Rob Spekkens for early contributions to this
work. Part of the research was carried out during a visit at the
Perimeter Institute, which N.H.L. and P.F.S. gratefully
acknowledge for the hospitality. During the completion of this
work, we learned about similar results obtained by Nicolas Gisin
and Sofyan Iblisdir
\cite{gi},
whom we kindly acknowledge for their correspondence. We also wish
to thank
 Emilio Bagan, Sofyan Iblisdir and Ramon Mu\~noz-Tapia
 for constructive criticism of an earlier
version of this paper.

Work by N.H.L was supported by a grant from the Technion Graduate
School. P.F.S is grateful to the EU (Grant HPRN-CT-2002-002777)
and to Prof. Joseph Avron for supporting this work.

\end{document}